\documentclass[12pt]{article}
\usepackage[dvips]{color}
\usepackage{epsfig}
\usepackage{amsmath}
\usepackage{cite}
\usepackage{graphicx}
\usepackage{color}
\usepackage{hyperref}
\usepackage{subfigure}

\begin{document}
\begin{center}
\Large{\bf The emergence of universal relations in the AdS black holes thermodynamics } \\
\small \vspace{1cm} {\bf J. Sadeghi$^{a}$\footnote {Email:~~~pouriya@ipm.ir}}, \quad
{\bf B. Pourhassan$^{b,c,d}$\footnote {Email:~~~b.pourhassan@du.ac.ir}}, \quad
{\bf S. Noori Gashti$^{a}$\footnote {Email:~~~saeed.noorigashti@stu.umz.ac.ir}}, \quad
{\bf S. Upadhyay$^{e,f,b}$\footnote {Email:~~~sudhakerupadhyay@gmail.com}\footnote{Visiting associate, Inter-University Centre for Astronomy and Astrophysics (IUCAA), Pune 411007, India.}},\quad {\bf  E. Naghd Mezerji$^{a}$ \footnote{Email:~~~e.n.mezerji@stu.umz.ac.ir}}\quad
\\
\vspace{0.5cm}$^{a}${Department of Physics, Faculty of Basic Sciences,\\
University of Mazandaran P. O. Box 47416-95447, Babolsar, Iran.}\\
\vspace{0.5cm}$^{b}${School of Physics, Damghan University, Damghan, 3671641167, Iran.}\\
\vspace{0.5cm}$^{c}${Physics Department, Istanbul Technical University, Istanbul 34469, Turkey.}\\
\vspace{0.5cm}$^{d}${Canadian Quantum Research Center 204-3002 32 Ave Vernon, BC V1T 2L7 Canada.}\\
\vspace{0.5cm}$^{e}${Department of Physics, K.L.S. College, Magadh University,  Nawada, Bihar 805110, India.}

\small \vspace{1cm}
\end{center}
\begin{abstract}
Our primary goal in this paper is to confirm new universal relations in black hole thermodynamics. We investigate the universal relations by selecting different black holes. First, we obtain the black holes' thermodynamic relations assuming a new minor correction is added to the AdS part of the action. Then we confirm the universal relations by performing a series of direct calculations. It is noteworthy that according to each of the properties related to black holes, a new universal relation can be obtained according to this method. We confirm two different types of these universal relations for various block holes. Furthermore, we also consider black holes in AdS space surrounded by perfect fluid. We use the small correction to the action and obtain the modified thermodynamic quantities. We achieve two new universal relations which correspond to the parameters of perfect fluid and magnetic charge of the Bardeen AdS Black Hole. Finally, the new universal relation leads us to understand the charge-to-mass ratio, i.e., WGC-like behavior. We also find that the weak gravity conjecture condition is satisfied for the black hole surrounded by perfect fluid.
\end{abstract}


\section{Literature Survey and Motivation}

The study of universal relations obtained from different methods is one of the most important points for earlier years which is considered by researchers. Physics is no exception. Recently, physicists have done many works in this context to be a beacon of hope for the unification of fundamental forces. Researchers have done much work to achieve universal relations in recent years. For example,  Davida and  Nian \cite{36} investigated the universal entropy and Hawking radiation of near-extremal $AdS_{4}$ black holes. Chen, Hong, and  Tao \cite{37} also studied universal thermodynamic extremality relations for charged AdS black holes surrounded by quintessence. In this paper, we want to study some universal relations between corrections to entropy and extremality bounds. Given the universal relations implications, universal thermodynamic extremality relations have not been fully explored in studies related to specific black holes. Therefore, in this work, we want to entirely evaluate several black holes with respect to this universal relation. Thus, by selecting several black holes in the AdS space, using a small constant correction added to the action, we obtain the modified thermodynamic relations and examine them to confirm the universal relations. The main point is using the new small correction in action for making modified thermodynamic relations that this correction somehow connects to the AdS part. So it leads to modifying parameters such as mass, temperature, etc., in AdS black holes. With respect to the first law of thermodynamics, we can establish universal relation between some of them that we fully explain in the next section, which is due to this small correction that is added to the AdS part of the action. Also, we investigate some new universal relations connected to weak gravity conjecture. Recently, this issue with some other conditions was examined we will mention. But in this paper, we want to challenge the new condition we pointed out earlier. As we said, one of the methods used in this theme, which is based on string theory, is the swampland program \cite{1}, which describes areas generally compatible with quantum gravity by stating some criteria. One of these criteria is the weak gravity conjecture {WGC), which has been widely studied in the last few years \cite{2}. This conjecture states that gravity is the weakest force, that is, there are states whose mass $(M)$ is less than their charge $(Q)$ which is $\frac{Q}{M}\geq1$ \cite{3,4,5,6,7,8,9,10,11,12,13,14,15,16,17}. Studies on this issue use black holes with different characteristics of mass, charge, etc. The important point is that this relation is not established for the black holes with naked singularity. Cosmic censorship is usually used to address this issue. One possible solution to this problem is to use correction to action that leads to modifying black hole solutions and, in a way, inverse the charge-to-mass ratio. For example, in Ref. \cite{18}, higher-derivative corrections are used for the mentioned issue. The result indicates that the charge-to-mass ratio is larger than $1$ using these higher-derivative operators and satisfies the universal relation. The idea of correcting higher derivatives has been widely used in the literature. One can find some works that have been done to limit effective theories using WGC in Refs. \cite{19,20,21,22,23,24,25,26,27,28,29,30,31}. Concepts related to WGC and obtaining universal relations have been explored in many works in recent years. In a recent investigation,  researchers prove the concept of WGC in flat space concerning Wald entropy \cite{22}. The concepts related to entropy changes due to the charge and mass of black holes and higher-derivative corrections have been studied to prove the notion of WGC about black holes. Goon and Penco\cite{32} also proposed a universal relation. This universal relation has also been studied for charged AdS black holes in four-dimensional space, which exhibits black hole WGC-like behavior \cite{33}. Other work has also been done on the Goon-Penco universal relation, for example, in examining four-derivative corrections competencies to the geometry of charged AdS black holes.
For further review of this relation, one can also look into Refs. \cite{34,35}.  Furthermore, In their previous work, the authors of this article also examined black holes surrounded by quintessence dark energy and cloud of string, which had beautiful results and led to the introduction of new universal relations \cite{040}. We want to examine another form of universal relation for different black holes. We add small constant corrections to the corresponding action and obtain the modified thermodynamic quantities and relations. These modified quantities help us to perform a series of calculations and confirm the universal relations. We also see the effect of small corrections on the thermodynamic relations. The corresponding correction increases the charge-to-mass ratio, which indicates    WGC-like behavior. To study thermodynamic relationships and the WGC, we select   Reissner-Nordst\"{o}m-AdS, rotating Bardeen and Kerr-Newman-AdS black holes surrounded by perfect fluid matter. As we know, dark matter also includes cold dark matter (CDM), warm dark matter (WDM), scalar background dark matter (SFDM), and perfect fluid dark matter (PFDM). Here, we combine PFDM with a black hole solution and investigate a particular type of universal relation as in Refs. \cite{040, 041,042,043,044,045,046,047,048,049,050,051,052,053,054,055,056,057,058,059}. Therefore, in addition to  PFDM,
we consider black holes with string fluid and obtain the new universal
relation as in Refs. \cite{060,061,062,063,064,065,066,067,068}.

The paper is organized as follows.   In Sections II, III, IV, V, VI, and VII, we discuss the universal relations of AdS Schwarzschild black holes, charged BTZ black holes, charged rotating BTZ black holes, accelerating black holes, charged accelerating black holes, and rotating accelerating black holes, respectively. In Section VIII, we also confirm the universal thermodynamic relations for the Reissner-Nordstr\"{o}m-AdS black hole with the perfect fluid dark matter due to a minor constant correction of the action with respect to confirming the WGC-like behavior.
  In Section IX, we consider Kerr-Newman-AdS black hole
surrounded by perfect fluid matter. Here also, we achieve the new universal relation and study the role of WGC on the corresponding system. In Section X, we consider rotating Bardeen black holes in AdS space surrounded by perfect fluid, and we confirm a new universal relation. Finally, we discuss and conclude in the last section.

\section{The Models}
In this section, we select several black holes. Since selected black holes have been investigated in the literature in different structures, various quantities such as their unmodified mass and temperature have been calculated that based on the references in each subsection, and can further study them.
 Therefore, in this article, we intend to calculate the modified structures of these quantities after applying a small correction that is added to the AdS part of the action of the corresponding black hole.
Then, by performing a series of mathematical calculations specified in the following, we create universal relationships between different structures according to the first law of thermodynamics.
The establishment of these universal relations in different formats, such as using higher derivatives, which we have already mentioned, has been investigated.
Now we want to show that this modification is minor and is added to the AdS part of the action; we can talk about universal relationships again, or is it possible to create a new relationship according to the thermodynamic components and their conjugates?
So, we selected several AdS black holes and will proceed with the calculations according to the mentioned process. Due to the modification of the action, the black hole solution is also modified. Therefore, some thermodynamic quantities of black holes will also be modified. Note that most of the time, one assumes the AdS radius l to be constant, although exciting consequences have been obtained by mapping it to thermodynamic pressure. Now we want to consider the small shift $\epsilon$ for the AdS radius and, with this change in this parameter, challenge the proof of the thermodynamic universal relations
\subsection{AdS Schwarzschild Black Holes}
Concerning all the concepts discussed above, we  examine the universal relations by considering different black holes.
In this context, we first apply on the AdS Schwarzschild black hole \cite{39}. The  action of AdS Schwarzschild  black hole is given by
\begin{equation}\label{1}
\mathcal{S}=\frac{1}{16\pi }\int d^{5}x\sqrt{-g}(R-2\Lambda),
\end{equation}
where the gravitational constant  $G=1$ (for simplicity)  and cosmological constant $\Lambda=-\frac{1}{l^{2}}$. The solution to the Einstein field equation for this block hole is written  as follows
\begin{equation}\label{2}
ds^{2}=f(r)dt^{2}+\frac{dr^{2}}{f(r)}+r^{2}d\Omega^{2},
\end{equation}
where  $f(r)$  has the following form:
\begin{equation}\label{3}
f(r)=1+\frac{r^{2}}{l^{2}}-\frac{2M_0}{r}.
\end{equation}
Here $M_0$  characterizes the mass of the black hole. The outer and inner horizons associated with a black hole are calculated from $f(r)=0$, and thus  the values of temperature, entropy, etc., can be easily calculated using the thermodynamic relations of the black hole.

 Now, we implement a minimal constant correction  by introducing $\epsilon$ to the  action as following:
\begin{equation}\label{4}
\mathcal{S}=\frac{1}{16\pi }\int d^{5}x\sqrt{-g}(R-(1+\epsilon)\times 2\Lambda).
\end{equation}
Due to the modification of the action, the black hole solution is also modified. Therefore, each of the thermodynamic values of the black holes will also be modified. Hence, the modified mass and temperature are obtained by considering a small constant correction $\epsilon$ and \cite{39,082} as follows,
\begin{eqnarray}
M&=&\frac{S}{8\pi}+\frac{S^{3}(1+\epsilon)}{128  l^{2}\pi^{3}},\label{5} \\
T&=&\frac{1}{8\pi}+\frac{3 S^{2}(1+\epsilon)}{128  l^{2}\pi^{3}}.\label{6}
\end{eqnarray}
 For such black holes, the externality boundary is also modified. So with $T=0$, the extremal entropy bounded by a small constant correction   is given by,
\begin{equation}\label{7}
S=\frac{4\pi l}{\sqrt{-3(1+\epsilon)}}.
\end{equation}
Here, upon  solving the temperature equation, we chose the real positive entropy. Also, by solving the equation (\ref{5}), the constant correction takes the following value
\begin{equation}\label{8}
\epsilon=\frac{128 l^{2} M \pi^{3}-16 l^{2} \pi^{2} S-S^{3}}{S^{3}}.
\end{equation}
The derivative of $\epsilon$ with respect to $S$is given by
\begin{equation}\label{9}
\frac{\partial\epsilon}{\partial S}=\frac{-16 l^{2}  \pi^{2}-3 S^{2} (1+\epsilon)}{S^{3}}.
\end{equation}
By combining the equations (\ref{6}), (\ref{7}) and (\ref{9}), we will have a relation at $M\rightarrow M_{ext}$ as follows,

\begin{equation}\label{10}
-T\frac{\partial S}{\partial\epsilon}=\frac{1}{6\sqrt{3} l^{2}}\left(-\frac{l^{2}}{(1+\epsilon)}\right)^{\frac{3}{2}}.
\end{equation}
After calculating the above relation, according to the equations (\ref{5}) and (\ref{7}), $M_{ext}$ is given by
\begin{equation}\label{11}
M_{ext}=\frac{1}{3\sqrt{3}}\left(-\frac{l^{2}}{(1+\epsilon)}\right)^{\frac{1}{2}}.
\end{equation}
The mass bound will increase with the constant correction parameter $\epsilon$, and in a way, this added correction can satisfy the conditions related to weak gravity conjecture. According to the concepts, we take the derivative of $M_{ext}$ with respect  to this constant parameter, we   have
\begin{equation}\label{12}
\frac{\partial M_{ext}}{\partial \epsilon}=\frac{1}{6\sqrt{3} l^{2}}\left(-\frac{l^{2}}{(1+\epsilon)}\right)^{\frac{3}{2}}.
\end{equation}
Incidentally, the two equations (\ref{10}) and (\ref{12}) are precisely  same. We first proved the Goon-Penco universal extremality relation for the AdS Schwarzschild black hole.

In the following, by selecting the other black holes and confirming the mentioned relation, we will examine the new universal relations that are somehow derived from the characteristics of black holes such as rotating.
\subsection{Charged BTZ Black Holes }
In this section, we  examine the universal relations  for charged BTZ black holes. The corresponding action  is given  as \cite{40}.
\begin{equation}\label{13}
\mathcal{S}=\frac{1}{16\pi}\int d^{3}x \sqrt{-g}(R-2\Lambda)
\end{equation}
The singular solution to the Einstein field equation for this black hole is
given by
\begin{equation}\label{14}
ds^{2}=-f(r)dt^{2}+f^{-1}(r)dr^{2}+r^{2} d\phi^{2},
\end{equation}
where $f(r)$ has the following form:
\begin{equation}\label{15}
f(r)=-M_0+\frac{r^{2}}{l^{2}}-\frac{Q^{2}}{2}\log(r).
\end{equation}
Here $M_0$ and $Q$ show the mass and charge of the black hole.
Following the above section, we introduce  a minimal constant parameter  $\epsilon$ to the  action (\ref{13}) as follows,
\begin{equation}\label{16}
\mathcal{S}=\frac{1}{16\pi}\int d^{3}x \sqrt{-g}(R-(1+\epsilon)\times 2\Lambda).
\end{equation}
This is obvious that the thermodynamics will also be modified due to
modification in action.  In this case, the modified mass and temperature are calculated concerning a small constant correction and with respect to \cite{40,082} as follows,
\begin{eqnarray}
M&=&\frac{S^{2}(1+\epsilon)}{16 \pi^{2} l^{2}}-\frac{Q^{2}}{2}\log\left[\frac{S}{4\pi}\right], \label{17}\\
T&=&-\frac{Q^{2}}{2S}+\frac{S(1+\epsilon)}{8l^{2}\pi^{2}}.\label{18}
\end{eqnarray}
For charged BTZ black hole, the externality boundary is also modified. So with $T=0$, the extremal entropy bound is obtained. After solving the temperature equation, we choose the real positive entropy, which is given by,
\begin{equation}\label{19}
S=\frac{2 l\pi Q}{\sqrt{1+\epsilon}}.
\end{equation}
Also, by simplifying the equation (\ref{17}), the constant correction parameter takes following value in this case:
\begin{equation}\label{20}
\epsilon=\frac{16l^{2}M\pi^{2}-S^{2}+8l^{2}\pi^{2}Q^{2}\log\left[\frac{S}{4\pi}\right]}{S^{2}}.
\end{equation}
The derivative of $\epsilon$ with respect to $S$ yields
\begin{equation}\label{21}
\frac{\partial\epsilon}{\partial S}=\frac{8l^{2}\pi^{2}Q^{2}-2S^{2}(1+\epsilon)}{S^{3}}.
\end{equation}
By combining the equations (\ref{18}), (\ref{19}) and (\ref{21}),  the mass takes $M\rightarrow M_{ext}$ as follows,
\begin{equation}\label{22}
-T\frac{\partial S}{\partial\epsilon}=\frac{Q^{2}}{4(1+\epsilon)},
\end{equation}
After calculating the above relation, according to the equations (\ref{17}) and (\ref{19}),   $M_{ext}$ takes following value:
\begin{equation}\label{23}
M_{ext}=\frac{Q^{2}}{4}\left(1-\log\left[\frac{l^{2}Q^{2}}{4(1+\epsilon)}\right]\right).
\end{equation}
This added correction can satisfy the conditions related to weak gravity conjecture. Upon taking   derivative with respect to the constant parameter, we will have
\begin{equation}\label{24}
\frac{\partial M_{ext}}{\partial \epsilon}=\frac{Q^{2}}{4(1+\epsilon)}.
\end{equation}
Here also,  equations (\ref{22}) and (\ref{24}) are precisely the same.
This justifies the Goon-Penco universal extremality relation for the charged BTZ black hole.
\subsection{Charged  Rotating BTZ Black Holes }
In this section, we chose the charged  rotating BTZ black holes \cite{43,44} in order to obtain the modified thermodynamic relations. We also consider the universal thermodynamic relations. First, we introduce the action and corresponding lapse function for the CR-BTZ black hole as follows,

\begin{equation*}\label{24}
\mathcal{S}=\frac{1}{2\pi}\int d^{3}x\sqrt{-g}\big(R+2\Lambda-\frac{\pi}{2}F_{\mu\nu}F^{\mu\nu}\big)
\end{equation*}

\begin{equation*}\label{24}
f(R)=-M+\frac{r^{2}}{l^{2}}+\frac{J^{2}}{4r^{2}}-\frac{\pi}{2}Q^{2}\ln r
\end{equation*}

Now we consider the small correction in action so that we will have,

\begin{equation*}\label{24}
\mathcal{S}=\frac{1}{2\pi}\int d^{3}x\sqrt{-g}\big(R+(1+\epsilon)\times2\Lambda-\frac{\pi}{2}F_{\mu\nu}F^{\mu\nu}\big)
\end{equation*}

Therefore, the modified thermodynamic parameters such as mass, temperature, angular velocity  with respect to constant small correction $\epsilon$ and \cite{40,43,44,082} as follows
\begin{eqnarray}
M&=&\frac{4 J^{2} \pi^{2}}{S^{2}}+\frac{S^{2}(\epsilon+1)}{16l^{2}\pi^{2}}-\frac{1}{2}\pi Q^{2} \log\left[\frac{S}{4\pi}\right],\label{48}\\
T &=&-\frac{8 J^{2} \pi^{2}}{S^{3}}-\frac{\pi Q^{2}}{2S}+\frac{S(1+\epsilon)}{8l^{2}\pi^{2}}, \label{49}\\
\Omega &=&\frac{8J\pi^{2}}{S^{2}}. \label{50}
\end{eqnarray}
Now  $T=0$ from equation (\ref{49}), the extremal entropy bound  can be obtained. After solving this temperature equation, we choose the real positive entropy as
\begin{equation}\label{51}
S=\sqrt{\frac{2 l^{2}\pi^{3}Q^{2}}{1+\epsilon}+\frac{2\pi^{2}l (16J^{2}+l^{2}\pi^{2}Q^{4}+16J^{2}\epsilon)^\frac{1}{2}}{1+\epsilon}}.
\end{equation}
From  equation (\ref{48}), we obtain
\begin{equation}\label{52}
\epsilon=\frac{-64S^{2}l^{2}\pi^{4}+16l^{2}M\pi^{2}S^{2}-S^{4}+8l^{2}\pi^{3}Q^{2}S^{2}}{S^{4}}.
\end{equation}
The derivative with respect to $S$ leads to
\begin{equation}\label{53}
\frac{\partial\epsilon}{\partial S}=\frac{128J^{2}l^{2}\pi^{4}+8l^{2}\pi^{3}S^{2}Q^{2}-2S^{4}(1+\epsilon)}{S^{5}}.
\end{equation}
Exploiting relations (\ref{49}), (\ref{51}) and (\ref{53}), we  have a relation at $M\rightarrow M_{ext}$ as follows
\begin{equation}\label{54}
-T\frac{\partial S}{\partial\epsilon}=\frac{l^{2}\pi Q^{2}+\sqrt{l^{4}\pi^{2}Q^{4}+16J^{2}l^{2}(1+\epsilon)}}{8l^{2}(1+\epsilon)}.
\end{equation}
The expression for $M_{ext}$ in this case is given as
\begin{equation}\label{55}
M_{ext}=\frac{\sqrt{l^{4}\pi^{2}Q^{4}+16J^{2}l^{2}(1+\epsilon)}}{4l^{2}}-\frac{\pi Q^{2}}{4}\log\left[\frac{l^{2} \pi Q^{2}+ \sqrt{l^{4}\pi^{2}Q^{4}+16J^{2}l^{2}(1+\epsilon)}}{8(1+\epsilon)}\right].
\end{equation}
The derivative of $M_{ext}$ with respect to  $\epsilon$ leads to
\begin{equation}\label{56}
\frac{\partial M_{ext}}{\partial \epsilon}=\frac{l^{2}\pi Q^{2}+\sqrt{l^{4}\pi^{2}Q^{4}+16J^{2}l^{2}(1+\epsilon)}}{8l^{2}(1+\epsilon)}.
\end{equation}
The equations (\ref{54}) and (\ref{56}) are the equal, so the universal relation for this black hole is also satisfied. The concepts presented and the rotating nature of this black hole lead us to examine another universal relation. We now study the new universal relation. Therefore, concerning equations (\ref{51}) and (\ref{55}), the shifting mass bound is given by
\begin{equation}\label{57}
\frac{\partial M_{ext}}{\partial \epsilon}=\frac{S^{2}}{16l^{2}\pi^{2}}.
\end{equation}
 From equation (\ref{52}), it is matter of calculation only to show
\begin{equation}\label{58}
\frac{\partial J}{\partial \epsilon}=-\frac{S^{4}}{128Jl^{2}\pi^{4}}.
\end{equation}
This expression further simplifies  to
\begin{equation}\label{59}
-\Omega \frac{\partial J}{\partial \epsilon} =\frac{S^{2}}{16l^{2}\pi^{2}}.
\end{equation}
As one can see, the two equations (\ref{57}) and (\ref{59}) are the same. This relation is due to the rotating nature of black holes  and, in this extremal limit, this equation is well proved.  We observed that a new universal relation could be defined for a black hole based on its characteristics. In fact, black holes with other features can be considered and examined for the new universal relations. It is also possible to introduce new universal thermodynamic relation concerning the specific characteristics of black holes.
\subsection{ Accelerating Black Holes }
In previous two sections, we proved the universal relation for the  Schwarzschild and charged BTZ  black holes. In this section, we consider an accelerating black hole \cite{41} to check the universal relation. The metric for the accelerating black hole  is given by
\begin{equation}\label{025}
ds^{2}=\frac{1}{\omega^{2}}\left[f(r)dt^{2}-f^{-1}(r)dr^{2}- r^{2}\frac{d\theta^{2}}{g(\theta)}+r^{2}g(\theta)\sin^{2}\theta\frac{d\phi^{2}}{K^{2}}
\right],
\end{equation}
where $K$ is the conical deficit and
\begin{equation}\label{026}
f(r)=(1- A^{2} r^{2})(1-\frac{2m}{r})+\frac{r^{2}}{l^{2}},\hspace{1cm} g(\theta)=1+2mA\cos\theta,\hspace{1cm}\omega=1+Ar\cos\theta.
\end{equation}
Here  $A$ is the acceleration and $m$ is  the mass scale of the black hole.
We can modify the action with the small correction constant parameter. So with respect to constant small correction $\epsilon$ and \cite{41}, each of the thermodynamic values of the black holes also gets modified. Hence, the modified mass and temperature are obtained as follows:
\begin{eqnarray}
M&=&\frac{S^{3}+l^{2}(16 \pi^{2} S(1+\epsilon)-A^{2} S^{3} (1+\epsilon))}{8l^{2}(16 \pi^{3}-A^{2} \pi S^{2})},\label{027}\\
T&=&\frac{48 \pi^{2} S^{2}-A^{2} S^{4}+l^{2}(-16 \pi^{2}+A^{2}S^{2})^{2}(1+\epsilon)}{8l^{2} \pi (-16 \pi^{2}+A^{2} S^{2})^{2}}. \label{028}
\end{eqnarray}
So with $T=0$, the extremal entropy bounded by a small constant correction   is given by
\begin{equation}\label{029}
S=\sqrt{\frac{-24 \pi^{2}+16A^{2}l^{2}\pi^{2}(1+\epsilon)-8\pi^{2}\sqrt{9-8A^{2}l^{2}(1+\epsilon)}}{-A^{2}+A^{4}l^{2}(1+\epsilon)}}.
\end{equation}
Here  chose the real positive entropy. After simplifying   equation (\ref{027}),we  get
\begin{equation}\label{030}
\epsilon=\frac{S^{3}-l^{2}(8M \pi -S)(16 \pi^{2} - A^{2}S^{2})}{l^{2}( -16 \pi^{2} S+A^{2} S^{3})}.
\end{equation}
Now, the derivative of $\epsilon$ with respect to $S$  yields
\begin{equation}\label{031}
\frac{\partial\epsilon}{\partial S}=\frac{-48 \pi^{2} S+A^{2} S^{3}}{l^{2}(-16\pi^{2}+A^{2}S^{2})^{2}}-\frac{(1+\epsilon)}{S}.
\end{equation}
Exploiting equations  (\ref{028}), (\ref{029}) and (\ref{031}), we   have a relation at $M\rightarrow M_{ext}$ as
\begin{equation}\label{032}
-T\frac{\partial S}{\partial\epsilon}=\sqrt{\frac{l^{2}(1+\epsilon)}{-6+4A^{2}l^{2}(1+\epsilon)+2\sqrt{9-8A^{2}l^{2}(1+\epsilon)}}}.
\end{equation}
According to the equations (\ref{027}) and (\ref{029}), $M_{ext}$ is given by
\begin{equation}\label{033}
M_{ext}=\frac{(-1+A^{2}l^{2}(1+\epsilon))(-3+\sqrt{9-8A^{2}l^{2}(1+\epsilon)})\sqrt{\frac{ (1+\epsilon)}{-6+4A^{2}l^{2}(1+\epsilon)+2\sqrt{9-8A^{2}l^{2}(1+\epsilon)}}}}{\sqrt{2}A^{2}l (-1+\sqrt{9-8A^{2}l^{2}(1+\epsilon)})}.
\end{equation}
This added correction can satisfy the conditions related to weak gravity conjecture.  We calculate the derivative of above relation according to this constant parameter $\epsilon$ and have
\begin{equation}\label{034}
\frac{\partial M_{ext}}{\partial \epsilon}=\sqrt{\frac{l^{2}(1+\epsilon)}{-6+4A^{2}l^{2}(1+\epsilon)+2\sqrt{9-8A^{2}l^{2}(1+\epsilon)}}}.
\end{equation}
The relations (\ref{032}) and (\ref{034}) coincide and the Goon-Penco universal extremality relation for this black hole is also justified.

\subsection{Charged Accelerating Black Holes }
 In this section we generalize the result for the charged accelerating black hole \cite{42}.  According to the above concepts in the previous section for the accelerating black hole, we add the charge to this black hole. By introducing  constant correction parameter $\epsilon$ and with respect to \cite{42}, the solution and thermodynamic parameters get modified. We obtain modified mass and temperature as follows
\begin{eqnarray}
M&=&\frac{2Q^{2}\pi}{S}+\frac{16l^{2} \pi^{2} S (1+\epsilon)+S^{3}-A^{2}l^{2}S^{3}(1+\epsilon)}{128l^{2} \pi^{3}- 8A^{2} l^{2} \pi S^{2}},\label{035}\\
 T&=&\frac{1+\epsilon}{8\pi}-\frac{2Q^{2}\pi}{S^{2}}+\frac{48 \pi^{2} S^{2} -A^{2}S^{4}}{8\pi l^{2}(-16 \pi^{2}+A^{2}S^{2})^{2}}.\label{036}
\end{eqnarray}
For vanishing temperature, the extremal entropy bound   is obtained. Four different values are obtained for entropy, two of which are imaginary, one is negative and the fourth is positive. After a very straightforward simplification and while maintaining the basic structure of entropy, it will be used in calculations related to proving the universal relation.   First by solving the equation (\ref{035}), we obtain
\begin{equation}\label{037}
\epsilon=\frac{16Q^{2} \pi^{2}(A^{2}-\frac{16\pi^{2}}{S^{2}})+M(\frac{128\pi^{3}}{S}-8A^{2}\pi S)-16\pi^{2}-\frac{S^{2}}{l^{2}}+A^{2}S^{2}}{16 \pi^{2}-A^{2}S^{2}}.
\end{equation}
By taking the derivative with respect to $S$ and concerning the positive value of entropy and implementing the equation (\ref{036}) along with   $M\rightarrow M_{ext}$, we  have
\begin{equation}\label{038}
-T\frac{\partial S}{\partial\epsilon}=\frac{A^{4}l^{2}Q^{2}\sqrt{\frac{3}{A^{2}}+\frac{l^{2}(-1+A^{2}Q^{2})}{A^{2}l^{2}-(1+\epsilon)}}}{2\sqrt{2}(A^{2}l^{2}-(1+\epsilon))(A^{4}l^{2}Q^{2}-(1+\epsilon))}.
\end{equation}
According to the equation  (\ref{035})  one can obtain
\begin{equation}\label{039}
\frac{\partial M_{ext}}{\partial \epsilon}=\frac{A^{4}l^{2}Q^{2}\sqrt{\frac{3}{A^{2}}+\frac{l^{2}(-1+A^{2}Q^{2})}{A^{2}l^{2}-(1+\epsilon)}}}{2\sqrt{2}(A^{2}l^{2}-(1+\epsilon))(A^{4}l^{2}Q^{2}-(1+\epsilon))}.
\end{equation}
In this case also,  the universal relation is satisfied as   equations (\ref{038}) and (\ref{039}) are the same.
\subsection{Rotating Accelerating Black Holes }

In this section, we select a rotating accelerating black hole \cite{42}. We follow the same procedure as in the  previous sections. The  point to be noted in this case is that the black hole under consideration is rotating.  By considering the small constant correction parameter and \cite{42}, we obtain each of the modified thermodynamic parameters such as mass, temperature, and angular velocity as follows,

\begin{eqnarray}
M&=&\frac{(16 J^{2} \pi^{2}+S^{2})(S^{2}+l^{2}(16\pi^{2}-A^{2}S^{2}(1+\epsilon))}{8l^{2}(16\pi^{3}S-A^{2}\pi S^{3})}\label{040}\\
T&=&\frac{A\times B}{C},\label{041}\\
\Omega &=&\frac{4J\pi^{2}(S^{2}+l^{2}(16\pi^{2}-A^{2}S^{2}(1+\epsilon)))}{l^{2}(16\pi^{3}S-A^{2}\pi S^{3})},\label{042}
\end{eqnarray}
where
\begin{eqnarray}
A&=&48 \pi^{2} S^{4}-A^{2}S^{6}+l^{2} S^{2}(256 \pi^{4}+A^{4} S^{4}(1+\epsilon)-16A^{2} \pi^{2} S^{2}(2+3\epsilon)),\\
B&=&-16J^{2} \pi^{2}(-S^{2}(16 \pi^{2}+A^{2} S^{2})+l^{2}(256 \pi^{4}+16 A^{2} \pi^{2} S^{2}(-2+\epsilon)\nonumber\\
&+&A^{4} S^{4}(1+\epsilon))),\\
C&=&8l^{2} \pi S^{2}(-16 \pi^{2}+A^{2} S^{2})^{2}.
\end{eqnarray}
 So concerning the equation (\ref{041}) that means $T=0$, the extremal entropy bound is obtained. Still, according to the calculations, four different values are obtained for entropy, two of which are imaginary, one is negative, and the fourth is positive. After a very straightforward simplification and while maintaining the basic structure of entropy, it will be used in calculations related to proving the universal relation. Hence, first, by solving the equation (\ref{040}), one can get
\begin{equation}\label{043}
\epsilon=-1+\frac{1}{l^{2}A^{2}}+\frac{8\pi M S}{16J^{2} \pi^{2}+S^{2}}+\frac{16\pi^{2}}{A^{2}}\left(\frac{16J^{2} \pi^{2}-8\pi M S+S^{2}}{16 J^{2} \pi^{2}S^{2}+S^{4}}\right).
\end{equation}
We take the derivative of  $S$  with respect to $\epsilon$ in (\ref{043}). By combining the equation (\ref{041}) and the positive entropy value as well as according to the condition $M\rightarrow M_{ext}$, we will have,
\begin{equation}\label{044}
-T\frac{\partial S}{\partial\epsilon}=\frac{2 J^{3} A^{2}(25+9Al (-2+Al(1+\sqrt{\epsilon}))(1+\sqrt{\epsilon}))}{3(-16 J^{2}A^{2}+9(-1+Al(1+\sqrt{\epsilon}))^{2})(-1+Al(1+\sqrt{\epsilon}))}.
\end{equation}
We consider the equation  (\ref{040}) and the  positive value of entropy  one can write
\begin{equation}\label{045}
\frac{\partial M_{ext}}{\partial \epsilon}=\frac{2 J^{3} A^{2}(25+9Al (-2+Al(1+\sqrt{\epsilon}))(1+\sqrt{\epsilon}))}{3(-16 J^{2}A^{2}+9(-1+Al(1+\sqrt{\epsilon}))^{2})(-1+Al(1+\sqrt{\epsilon}))}.
\end{equation}
Here, we observe that
  equations (\ref{044}) and (\ref{045}) are exactly same and the universal relation for this black hole are also satisfied. As mentioned above, the rotating nature of this black hole leads us to examine another universal relation. According to the concepts presented, we now study the new universal relation. Therefore, with respect to equation  (\ref{040})  and mass bound with respect to the condition, $M\rightarrow M_{ext}$ and solve the entropy in terms of $J$, we can obtain the new relation with regard to $\epsilon$ derivative as follows,
\begin{equation}\label{046}
\frac{\partial M_{ext}}{\partial \epsilon}=\frac{A^{2}(16J^{2}\pi^{2}+S^{2})^{2}(-S^{2}+l^{2}(-16 \pi^{2}+A^{2} S^{2}(1+\epsilon)))}{64 l^{2}M\pi^{2}(-16 \pi^{2}+A^{2}S^{2})^{2}}.
\end{equation}
Now considering the equations (\ref{042}) and (\ref{043}) we   have at $M\rightarrow M_{ext}$,
\begin{equation}\label{047}
-\Omega\left(\frac{\partial J}{\partial \epsilon}\right)=\frac{A^{2}(16J^{2}\pi^{2}+S^{2})^{2}(-S^{2}+l^{2}(-16 \pi^{2}+A^{2} S^{2}(1+\epsilon)))}{64 l^{2}M\pi^{2}(-16 \pi^{2}+A^{2}S^{2})^{2}}.
\end{equation}

Comparing the   equations (\ref{046}) and (\ref{047}), one can see that these two equations are the same. This relation is due to the rotating nature of black holes, and in this extremal limit, these equations are well confirmed.

In the next section, we will study both of these universal relations examined in this section for a charged rotating BTZ black hole in AdS space-time. Of course, the interesting point is that such a new universal relation can be confirmed by considering each black holes specific feature.

\subsection{Reissner-Nordstr\"{o}m AdS black hole with PFDM}
In this section, we  wish to study  the Goon-Penco universal extremality relation for Reissner-Nordstr\"{o}m AdS black hole with PFDM and  obtain the other new universal relation. So, we consider  Reissner-Nordstr\"{o}m AdS black hole. We use a small correction $\epsilon$  to the action, and obtain  the modified thermodynamic quantities and relations.  Here we first write the action which is given by \cite{068},
\begin{equation}\label{048}
\mathcal{S}=\int d^{4}x\sqrt{-g}\left[\frac{R}{16\pi G}-\frac{\Lambda}{8\pi G}+\frac{1}{4}F^{\mu\nu}F_{\mu\nu}+L_{DM}\right],
\end{equation}
where $G$, $\Lambda$, $F_{\mu\nu}$ and $L_{DM}$ are the Newton gravity constant, cosmological constant,
tensor of electromagnetic field and dark matter Lagrangian density, respectively. So, the solution of action with  PFDM  is as  following \cite{069},
\begin{equation}\label{049}
ds^{2}=-f(r)dt^{2}+f^{-1}(r)dr^{2}+r^{2}(d\theta^{2}+\sin^{2}\theta d\phi^{2}),
\end{equation}
where $f(r)$ is
\begin{equation}\label{050}
\begin{split}
f(r)=1-\frac{2M}{r}+\frac{Q^{2}}{r^{2}}+\frac{1}{3}\Lambda r^{2}+\frac{\alpha}{r}\ln\left(\frac{r}{|\alpha|}\right).
\end{split}
\end{equation}
Here, the $M$, $Q$ and $\alpha$ are the mass, charge and the intensity of the PFDM, respectively. $f(r)=0$ determine the outer and inner horizons. The thermodynamic relation such as temperature, entropy, etc can be easily calculated. The mass and temperature of Reissner-Nordstr\"{o}m AdS black hole with PFDM are given by,
\begin{equation}\label{051}
M=\frac{\sqrt{\frac{\pi}{2}}Q^{2}}{\sqrt{S}}+\frac{\sqrt{S}}{2\sqrt{2\pi}}-\frac{S^{\frac{3}{2}}}{4\sqrt{2}l^{2}
\pi^{\frac{3}{2}}}+\frac{1}{2}\alpha\log\left(\frac{\sqrt{S}}{\sqrt{2\pi}\alpha}
\right),
\end{equation}
and
\begin{equation}\label{052}
T=-\frac{\sqrt{\frac{\pi}{2}}Q^{2}}{\sqrt{S}}+\frac{1}{4\sqrt{2\pi S}}-\frac{3\sqrt{S}}{8\sqrt{2}l^{2}\pi^{\frac{3}{2}}}+\frac{\alpha}{4S}.
\end{equation}
Here, we consider a small constant correction as $\epsilon$ to the action. So, we have following modified action:
\begin{equation}\label{053}
\mathcal{S}=\int d^{4}x\sqrt{-g}\left[\frac{R}{16\pi G}+(1+\epsilon) (-\frac{\Lambda}{8\pi G})+\frac{1}{4}F^{\mu\nu}F_{\mu\nu}+L_{DM}\right].
\end{equation}
With respect to small constant correction $\epsilon$ which is added to action and \cite{068}, the modified mass and temperature are given, respectively, by
\begin{equation}\label{054}
M=\frac{\sqrt{\frac{\pi}{2}}Q^{2}}{\sqrt{S}}+\frac{\sqrt{S}}{2\sqrt{2\pi}}-\frac{S^{\frac{3}{2}}
(1+\epsilon)}{4\sqrt{2}l^{2}\pi^{\frac{3}{2}}}+\frac{1}{2}\alpha\log\left(\frac{\sqrt{S}}{\sqrt{2\pi}\alpha}\right),
\end{equation}
and
\begin{equation}\label{055}
T=-\frac{\sqrt{\frac{\pi}{2}}Q^{2}}{\sqrt{S}}+\frac{1}{4\sqrt{2\pi S}}-\frac{3\sqrt{S}(1+\epsilon)}{8\sqrt{2}l^{2}\pi^{\frac{3}{2}}}+\frac{\alpha}{4S}.
\end{equation}
We see here that the modified mass and temperature of the black hole are written in terms of entropy $S$, charge $Q$, $\alpha$ and correction parameter $\epsilon$. By solving equation (\ref{054}), we obtain constant correction parameter $\epsilon$ as
 \begin{equation}\label{056}
\epsilon=\frac{4\sqrt{2}l^{2}\pi^{2}Q^{2}-8l^{2}M\pi^{\frac{3}{2}}\sqrt{S}+2\sqrt{2}l^{2}S\pi-\sqrt{2}S^{2}+4l^{2}\sqrt{S}\alpha \pi^{\frac{3}{2}}\log(\frac{\sqrt{S}}{\sqrt{2\pi}\alpha})}{\sqrt{2}S^{2}}.
\end{equation}
 Now, we take derivative with respect to $S$, which yields
\begin{equation}\label{057}
\frac{\partial \epsilon}{\partial S}=\frac{l^{2}\pi(-8\pi Q^{2}-2S+\sqrt{2\pi S}(6M+\alpha-3\alpha\log(\frac{\sqrt{S}}{\sqrt{2\pi}\alpha})}{S^{3}}.
\end{equation}
We use  the equations (\ref{055}) and (\ref{057}) and then the corresponding limit is taken, which  simplifies equation  as follows
\begin{equation}\label{058}
-T\frac{\partial S}{\partial \epsilon}=-\frac{S^{\frac{3}{2}}}{4\sqrt{2}l^{2}\pi^{\frac{3}{2}}}.
\end{equation}
To obtain the second part of universal relation, we solve  the temperature equation and obtain the corresponding entropy. We use equations (\ref{054}),
 (\ref{055}) and corresponding entropy, to obtain
\begin{equation}\label{059}
\frac{\partial M_{ext}}{\partial\epsilon}= -\frac{S^{\frac{3}{2}}}{4\sqrt{2}l^{2}\pi^{\frac{3}{2}}}.
\end{equation}
We see that the equations (\ref{058}) and (\ref{059}) are exactly  same.  So, we first confirm   the Goon-Penco universal extremality relation for Reissner-Nordstr\"{o}m AdS black hole with PFDM.

In the second step, we  investigate another universal relation. So, we use the relation (\ref{056}) and calculate
 \begin{equation}\label{60}
\frac{\partial \epsilon}{\partial Q}=\frac{8l^{2}\pi^{2}Q}{S^{2}}.
\end{equation}
 By considering equation (\ref{60}), electric potential $\Phi=\frac{\sqrt{2\pi}Q}{\sqrt{S}}$ and extermality bound, we get
\begin{equation}\label{61}
-\Phi\frac{\partial Q}{\partial \epsilon}=-\frac{S^{\frac{3}{2}}}{4\sqrt{2}l^{2}\pi^{\frac{3}{2}}}.
\end{equation}
Here, also two equations (\ref{61}) and (\ref{059}) are same.

In the following, we   try to find  another universal relationship between mass and pressure ($P=\frac{3}{8\pi l^{2}}=-\frac{\Lambda}{8\pi}$). So with respect to equation (\ref{054}) we will have
\begin{equation}\label{62}
\frac{\partial P}{\partial \epsilon}=\frac{4\sqrt{2}P^{2}S^{2}}{-\frac{3S^{2}(1+\epsilon)}{\sqrt{2}l^{2}\pi}}.
\end{equation}
Therefore, according to $V=-\frac{1}{3}\sqrt{\frac{2}{\pi}}S^{\frac{3}{2}}(1+\epsilon)$ and extremal bound, we have following expression,
\begin{equation}\label{63}
-V\frac{\partial P}{\partial \epsilon}=-\frac{S^{\frac{3}{2}}}{4\sqrt{2}l^{2}\pi^{\frac{3}{2}}}
\end{equation}
Two equations (\ref{61}) and (\ref{63}) are exactly  same and universal relation is proved. Now our goal is to get a new universal relation, the meaning of being new is to be related to a newly introduced parameter such as perfect fluid $(\alpha)$. To achieve such new universal relation, we use (\ref{056}) and obtain the following equation:
\begin{equation}\label{64}
\frac{\partial \epsilon}{\partial \alpha}=\frac{-4l^{2}\sqrt{S}\pi^{\frac{3}{2}}+4l^{2}\sqrt{S}\pi^{\frac{3}{2}}\log(\frac{\sqrt{S}}{\alpha\sqrt{2\pi}})}{\sqrt{2}S^{2}}.
\end{equation}
 Hence, by using equation (\ref{64}) and expression $\xi=-\frac{1}{2}+\frac{1}{2}\log(\frac{\sqrt{S}}{\alpha\sqrt{2\pi}})$ which is the conjugate the parameter $\alpha$, one can obtain
\begin{equation}\label{65}
-\xi\frac{\partial \alpha}{\partial \epsilon}=-\frac{S^{\frac{3}{2}}}{4\sqrt{2}l^{2}\pi^{\frac{3}{2}}}.
\end{equation}
We see here  two equations (\ref{65}) and (\ref{63}) are exactly  same. In fact, we obtained a new universal relation for the Reissner-Nordstr$\ddot{o}$m AdS black hole with PFDM, so this relationship is confirmed correctly.
\begin{figure}[h!]
 \begin{center}
 \subfigure[]{
 \includegraphics[height=5cm,width=5cm]{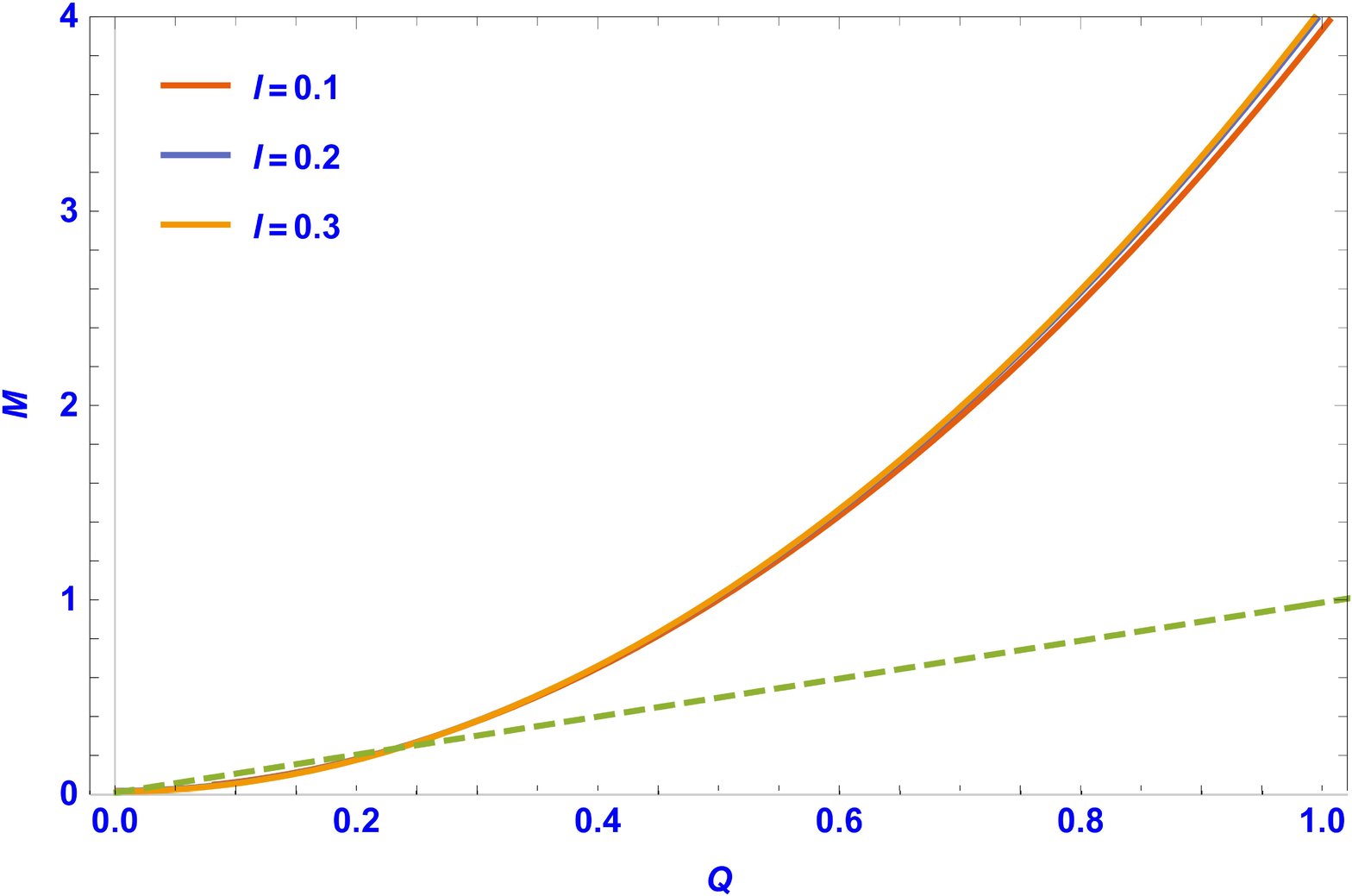}
 \label{1a}}
 \subfigure[]{
 \includegraphics[height=5cm,width=5cm]{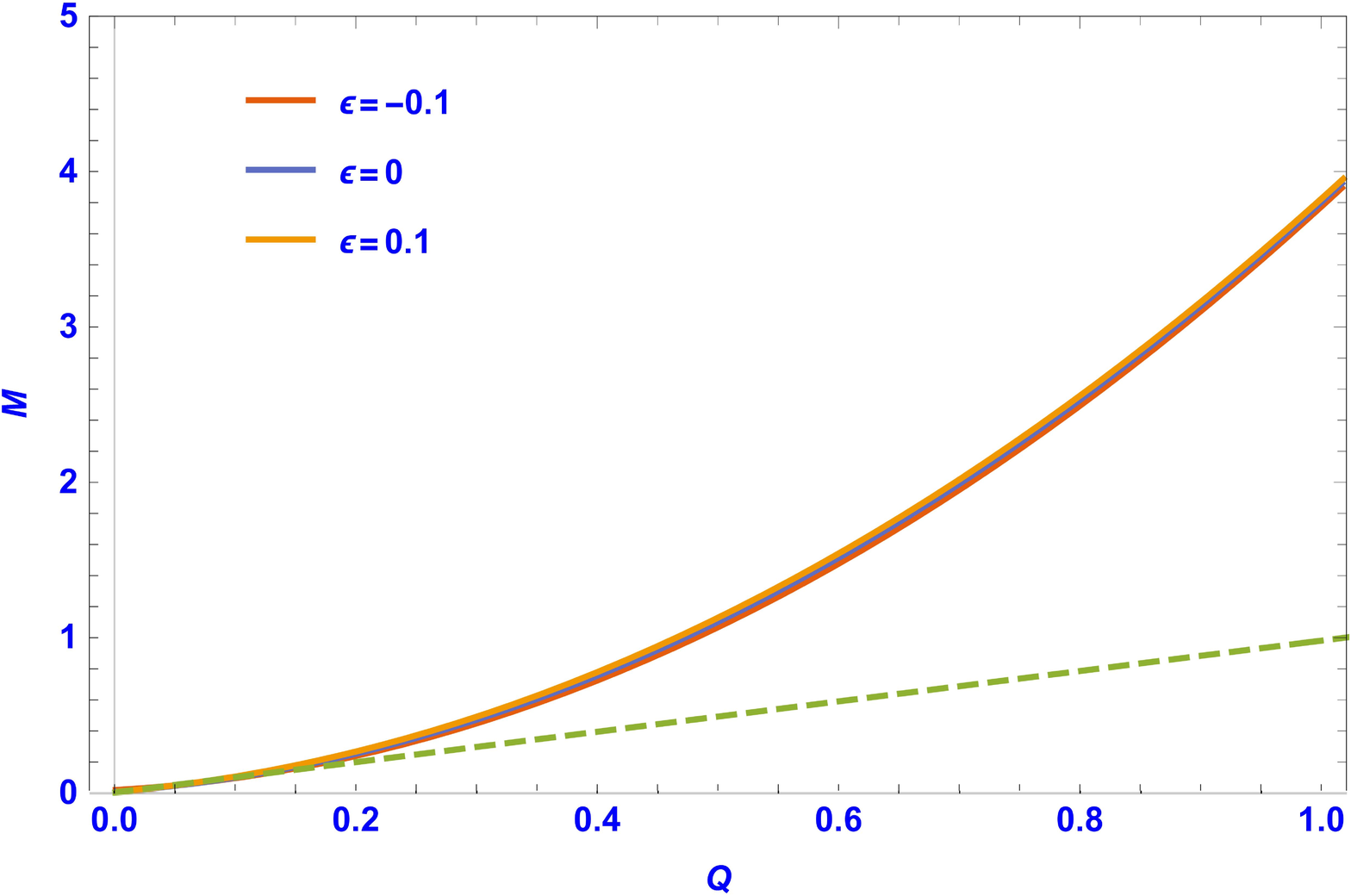}
 \label{1b}}
  \caption{\small{The plot of unmodified $M$ in term of $Q$ with respect to $l=0.1,0.2,0.3$ in fig. (a) and the plot of modified $M$ in term of $Q$ with respect to $l=0.1$ and $\epsilon=-0.1,0,0.1$ in fig. (b) }}
 \label{fig1}
 \end{center}
 \end{figure}
Now we draw some figures and compare the mass charge ratio and also  thermodynamic quantities in the modified and unmodified modes. We fix some  parameters  and draw figure of  the mass-to-charge, we have three cases as charge-mass ratio be bigger, equal and smaller than one. So, when we have extremality  the charge-mass ratio is one which is shown in the figure \ref{fig1}   by dashed line. For different radius $l$ of AdS in the uncorrected mass mode, the mass-to-charge ratio is greater than one as shown in the figure \ref{1a}. For the modified case we have figure \ref{1b}, when the constant correction is always a negative value, the mass of the black hole decreases, and when the small correction is positive, the mass also increases. Therefore, when we consider small negative correction, the charge-to-mass ratio always increases. As we mentioned in the text, it can behave like  WGC. So, we see that the small correction and its parameter in the metric background play an important role  for the existence of WGC.

 In the next section, we calculate same universal relation for Kerr-Newman AdS black hole surrounded by perfect fluid matter and also obtain a new universal relation. Then we compare the corresponding results  of two section with respect to each other.

\subsection{Kerr-Newman AdS black hole surrounded by perfect fluid matter}
 In this section, we want to study the universal relations for Kerr-Newman AdS
 black hole surrounded by perfect fluid matter in Rastall gravity \cite{070}. The
 concepts of general relativity and its modification have always been discussed in
 various branch of  physics. Some researchers use it's modified form in concepts
 of conservation condition of
the energy-momentum tensor. One of these modified theories was also introduced by
Rastall. Rastall gravity is based on the hypothesis   introduced by $T^{\mu
\nu}_{;\mu}=\lambda R^{,\nu}$, where  $T_{\mu\nu}$ and $\lambda$ are
energy-momentum tensor and the Rastall parameter, respectively \cite{071,072}. The Kerr-Newman AdS black hole solution is
given by \cite{071}
\begin{eqnarray}
ds^{2}&=&\frac{\Sigma^{2}}{f(r)}dr^{2}+\frac{\Sigma^{2}}{f(\theta)}d\theta^{2}+
\frac{f(\theta)\sin^{2}\theta}{\Sigma^{2}}\left(a\frac{dt}{\Xi}-(r^{2}+a^{2})\frac{d
\phi}{\Xi}\right)^{2}-\frac{f(r)}{\Sigma^{2}}\left(\frac{dt}{\Xi}-a\sin^{2}\frac{d\phi}
{\Xi}\right)^{2},\label{66}\\
 f(r)&=&r^{2}-2Mr+a^{2}+Q^{2}-\frac{\Lambda}{3}r^{2}(r^{2}+a^{2})-\alpha
r^{\frac{1-3\omega}{1-3\kappa \lambda(1+\omega)}}\nonumber\\
f(\theta)&=&1+\frac{\Lambda}{3}a^{2}\cos^{2}\theta,\hspace{2cm}\Xi=1+\frac{\Lambda}
{3}a^{2},
\label{67}
\end{eqnarray}
  where  $M$, $Q$, $a$, $\alpha$ and $\lambda$ are the mass, charge, rotational
parameter, perfect fluid parameter and Rastall parameter, respectively. Likewise
the previous sections the outer and inner horizons are calculated by $f(r)=0$. With
respect to entropy of black hole, the thermodynamic quantities of Kerr-Newman AdS
black hole surrounded by perfect fluid such as mass, temperature and angular
velocity are given by
\begin{equation}\label{68}
 M= \frac{a^{2}\sqrt{\pi}}{\sqrt{S}}\frac{\sqrt{\pi}Q^{2}}{\sqrt{S}}+\frac{\sqrt{S}}{4\sqrt{\pi}}+\frac{a^{2}\sqrt{S}}{4l^{2}\sqrt{\pi}}
+\frac{S^{\frac{3}{2}}}{16l^{2}\pi^{\frac{3}{2}}}-2^{\frac{1-3\omega}{-1+3\kappa\lambda(1+\omega)}}\\
 \pi^{\frac{3(\kappa\lambda+(-1+\kappa\lambda)\omega)}{-2+6\kappa\lambda(1+\omega)}}
S^{-\frac{3(\kappa\lambda+(-1+\kappa\lambda)\omega)}{-2+6\kappa\lambda(1+\omega)}}\alpha,
 \end{equation}
\begin{eqnarray}\label{69}
 T&=&-\frac{a^{2}\sqrt{\pi}}{\sqrt{S}}-\frac{\sqrt{\pi}Q^{2}}{2S^{\frac{3}{2}}}+\frac{1}{8\sqrt{\pi S}}+\frac{a^{2}}{8l^{2}\sqrt{\pi S}}+\frac{3\sqrt{S}}{32l^{2}\pi^{\frac{3}{2}}}\\
&+&\frac{3\times2^{\frac{1-3\omega}{-1+3\kappa\lambda(1+\omega)}}
\times\pi^{\frac{3(\kappa\lambda+(-1+\kappa\lambda)\omega)}{-2+6\kappa\lambda(1+\omega)}}
S^{-1-\frac{3(\kappa\lambda+(-1+\kappa\lambda)\omega)}{-2+6\kappa\lambda(1+\omega)}}
\alpha(\kappa\lambda+(-1+\kappa\lambda)\omega))}{-2+6\kappa\lambda(1+\omega)},
\end{eqnarray}
and
\begin{equation}\label{70}
\Omega=\frac{2a\sqrt{\pi}}{\sqrt{S}}+\frac{a\sqrt{S}}{2l^{2}\sqrt{\pi}}.
\end{equation}
Now, by considering the small constant correction $\epsilon$ for Kerr-Newman AdS black hole surrounded perfect fluid and \cite{070}, we calculate the modified thermodynamic quantities as mass, temperature and angular velocity are given by
\begin{eqnarray}
M&=& \frac{a^{2}\sqrt{\pi}}{\sqrt{S}}\frac{\sqrt{\pi}Q^{2}}{\sqrt{S}}+\frac{\sqrt{S}}{4\sqrt{\pi}}+\frac{a^{2}\sqrt{S}(1+\epsilon)}{4l^{2}\sqrt{\pi}}
+\frac{(1+\epsilon)S^{\frac{3}{2}}}{16l^{2}\pi^{\frac{3}{2}}}\nonumber\\
&-&2^{\frac{1-3\omega}{-1+3\kappa\lambda(1+\omega)}}
 \pi^{\frac{3(\kappa\lambda+(-1+\kappa\lambda)\omega)}{-2+6\kappa\lambda(1+\omega)}}
S^{-\frac{3(\kappa\lambda+(-1+\kappa\lambda)\omega)}{-2+6\kappa\lambda(1+\omega)}}\alpha, \label{71}\\
T&=&-\frac{a^{2}\sqrt{\pi}}{\sqrt{S}}-\frac{\sqrt{\pi}Q^{2}}{2S^{\frac{3}{2}}}+\frac{1}{8\sqrt{\pi S}}+\frac{a^{2}(1+\epsilon)}{8l^{2}\sqrt{\pi S}}+\frac{3\sqrt{S}(1+\epsilon)}{32l^{2}\pi^{\frac{3}{2}}}\nonumber\\
&+&\frac{3\times2^{\frac{1-3\omega}{-1+3\kappa\lambda(1+\omega)}}
\times\pi^{\frac{3(\kappa\lambda+(-1+\kappa\lambda)\omega)}{-2+6\kappa\lambda(1+\omega)}}
S^{-1-\frac{3(\kappa\lambda+(-1+\kappa\lambda)\omega)}{-2+6\kappa\lambda(1+\omega)}}
\alpha(\kappa\lambda+(-1+\kappa\lambda)\omega))}{-2+6\kappa\lambda(1+\omega)},
  \label{72}
  \end{eqnarray}
 and
\begin{equation}\label{73}
\Omega=\frac{2a\sqrt{\pi}}{\sqrt{S}}+\frac{a\sqrt{S}(1+\epsilon)}{2l^{2.}\sqrt{\pi}}
\end{equation}
Here we note that  thermodynamic relation  lead us to consider   $\alpha$ as a conjugate to $\eta=-2^{\frac{1-3\omega}{-1+3\kappa\lambda(1+\omega)}}\\
\times\pi^{\frac{3(\kappa\lambda+(-1+\kappa\lambda)\omega)}{-2+6\kappa\lambda(1+\omega)}}
S^{-\frac{3(\kappa\lambda+(-1+\kappa\lambda)\omega)}{-2+6\kappa\lambda(1+\omega)}}\alpha$, charge  conjugate to electric potential as  $\Phi=\frac{2\sqrt{\pi}Q}{\sqrt{S}}$
and volume $V= \frac{2}{3}a^{2}\sqrt{S\pi+\frac{S^{\frac{3}{2}}}{6\sqrt{\pi}}} (1+\epsilon)$ conjugate to $P=-\frac{\Lambda}{8\pi}$. We use  such conjugate quantities and  prove   universal relations. On the other hand, the changes of mass and entropy in terms of small constant corrections $\epsilon$ can actually be a clue to the WGC. Therefore, by solving equation (\ref{71}), the constant correction parameter $ \epsilon $ is calculated by  \begin{eqnarray}
 \epsilon &=& -1+\frac{1}{\frac{a^{2}\sqrt{S}}{4l^{2}\sqrt{\pi}}+\frac{S^{\frac{3}{2}}}{16l^{2}\pi^{\frac{3}{2}}}}
 \left(M-\frac{a^{2}\sqrt{\pi}}{\sqrt{S}}-\frac{\sqrt{\pi}Q^{2}}{\sqrt{S}}-\frac{\sqrt{S}}{4\sqrt{\pi}}\right.\nonumber\\
 &+& \left. 2^{\frac{1-3\omega}{-1+3\kappa\lambda(1+\omega)}}
\times\pi^{\frac{3(\kappa\lambda+(-1+\kappa\lambda)\omega)}{-2+6\kappa\lambda(1+\omega)}}S^{-\frac{3(\kappa\lambda+(-1+\kappa\lambda)\omega)}{-2+6\kappa\lambda(1+\omega)}}\alpha\right).\label{74}
\end{eqnarray}
  Then, we take derivative with respect to $S$ and get
\begin{equation}\label{75}
\begin{split}
&\frac{\partial\epsilon}{\partial S}=\frac{A+B}{C},\\
&A=-\left(\frac{a^{2}}{8l^{2}\sqrt{\pi S}}+\frac{3\sqrt{S}}{32l^{2}\pi^{\frac{3}{2}}}\right)(1+\epsilon)+\frac{a^{2}\sqrt{\pi}}{2S^{\frac{3}{2}}}+\frac{\sqrt{\pi}Q^{2}}{2S^{\frac{3}{2}}}-\frac{1}{8\sqrt{\pi S}},\\
&B=-\frac{3\times2^{\frac{1-3\omega}{-1+3\kappa\lambda(1+\omega)}}
\times\pi^{\frac{3(\kappa\lambda+(-1+\kappa\lambda)\omega)}{-2+6\kappa\lambda(1+\omega)}}S^{-1-\frac{3(\kappa\lambda+(-1+\kappa\lambda)\omega)}{-2+6\kappa\lambda(1+\omega)}}\alpha(\kappa\lambda+(-1+\kappa\lambda)\omega))}{-2+6\kappa\lambda(1+\omega)},\\
&C= \frac{a^{2}\sqrt{S}}{4l^{2}\sqrt{\pi}}+\frac{S^{\frac{3}{2}}}{16l^{2}\pi^{\frac{3}{2}}}.
\end{split}
\end{equation}
Now by combining  two equations (\ref{72}) and (\ref{75}),   we have
\begin{equation}\label{76}
-T\frac{\partial S}{\partial \epsilon}=\frac{\sqrt{S}(4a^{2}\pi+S)}{16l^{2}\pi^{\frac{3}{2}}}.
\end{equation}
To get the second part of universal relation, we set $ T = 0 $. By solving the temperature, obtaining the entropy and  using equations (\ref{71}) and (\ref{72}), we obtain following equation:
\begin{equation}\label{77}
\frac{\partial M_{ext}}{\partial\epsilon}=\frac{\sqrt{S}(4a^{2}\pi+S)}{16l^{2}\pi^{\frac{3}{2}}}.
\end{equation}
Here we see that  two equations (\ref{76}) and (\ref{77}) are exactly  same. So, we confirmed  the Goon-Penco universal extremality relation.

 To investigate another universal relation, we use  the  equation  (\ref{74}) to obtain following relation:
 \begin{equation}\label{78}
\frac{\partial \epsilon}{\partial Q}=-\frac{2\sqrt{\pi}Q}{\sqrt{S}(\frac{a^{2}\sqrt{S}}{4l^{2}\sqrt{\pi}}+\frac{S^{\frac{3}{2}}}{16l^{2}\pi^{\frac{3}{2}}})}.
\end{equation}
From (\ref{78}) and  electric potential $\Phi$ as well as extermality bound,  we have,
\begin{equation}\label{79}
-\Phi\frac{\partial Q}{\partial \epsilon}=\frac{\sqrt{S}(4a^{2}\pi+S)}{16l^{2}\pi^{\frac{3}{2}}}.
\end{equation}
Also here, we see that the equation (\ref{79}) and (\ref{77}) are same, so the universal relation is also proved. Therefore,  with respect to pressure $P=\frac{3}{8\pi l^{2}}=-\frac{\Lambda}{8\pi}$ and equation (\ref{74}), we have
\begin{equation}\label{80}
\frac{\partial P}{\partial \epsilon}=-\frac{P^{2}(\frac{2}{3}a^{2}\sqrt{\pi S}+\frac{S^{\frac{3}{2}}}{6\sqrt{\pi}})}{\frac{\sqrt{S}(4a^{2}\pi+S)(1+\epsilon)}{16l^{2}\pi^{\frac{3}{2}}}}.
\end{equation}
The above equation and extremal bound lead us to  obtain following equation:
\begin{equation}\label{81}
-V\frac{\partial P}{\partial \epsilon}=\frac{\sqrt{S}(4a^{2}\pi+S)}{16l^{2}\pi^{\frac{3}{2}}}.
\end{equation}
Here also we see that  two equations (\ref{81}) and (\ref{77}) are exactly same.

 We also study the other universal relation, hence according to relation (\ref{74}) and  using the equation (\ref{73}) one can obtain
\begin{equation}\label{82}
-\Omega\frac{\partial a}{\partial \epsilon}=\frac{\sqrt{S}(4a^{2}\pi+S)}{16l^{2}\pi^{\frac{3}{2}}}.
\end{equation}
We see, two equations (\ref{82}) and (\ref{77}) are exactly  same. Now we will study the new universal relation with respect to $\eta$ which is conjugate to the perfect fluid parameter $\alpha$. So, according to equation (\ref{74}), we have
\begin{equation}\label{83}
\frac{\partial \epsilon}{\partial\alpha}=\frac{2^{\frac{1-3\omega}{-1+3\kappa\lambda(1+\omega)}}\times\pi^{\frac{3(\kappa\lambda+(-1+\kappa\lambda)\omega)}{-2+6\kappa\lambda(1+\omega)}}S^{-\frac{3(\kappa\lambda+(-1+\kappa\lambda)\omega)}{-2+6\kappa\lambda(1+\omega)}}}{(\frac{a^{2}\sqrt{S}}{4l^{2}\sqrt{\pi}}+\frac{S^{\frac{3}{2}}}{16l^{2}\pi^{\frac{3}{2}}})}.
\end{equation}
Now by using the $\eta$ and the equation (\ref{83}), we confirmed another universal relation which is calculated by,
\begin{equation}\label{84}
-\eta\frac{\partial \alpha}{\partial\epsilon}=\frac{\sqrt{S}(4a^{2}\pi+S)}{16l^{2}\pi^{\frac{3}{2}}}=\frac{\partial M_{ext}}{\partial \epsilon}.
\end{equation}
\begin{figure}[h!]
\begin{center}
\subfigure[]{
\includegraphics[height=5cm,width=5cm]{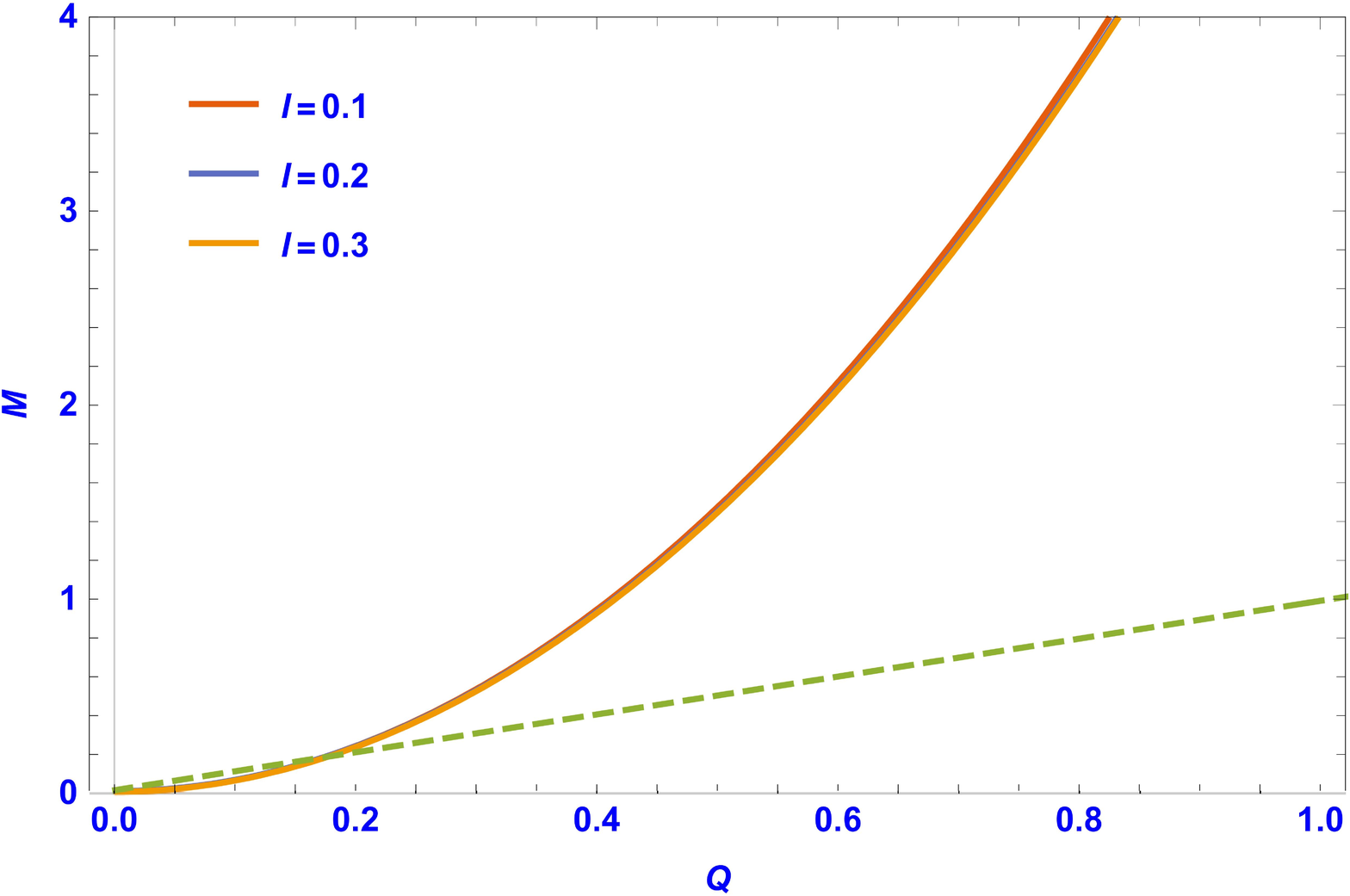}
\label{2a}}
\subfigure[]{
\includegraphics[height=5cm,width=5cm]{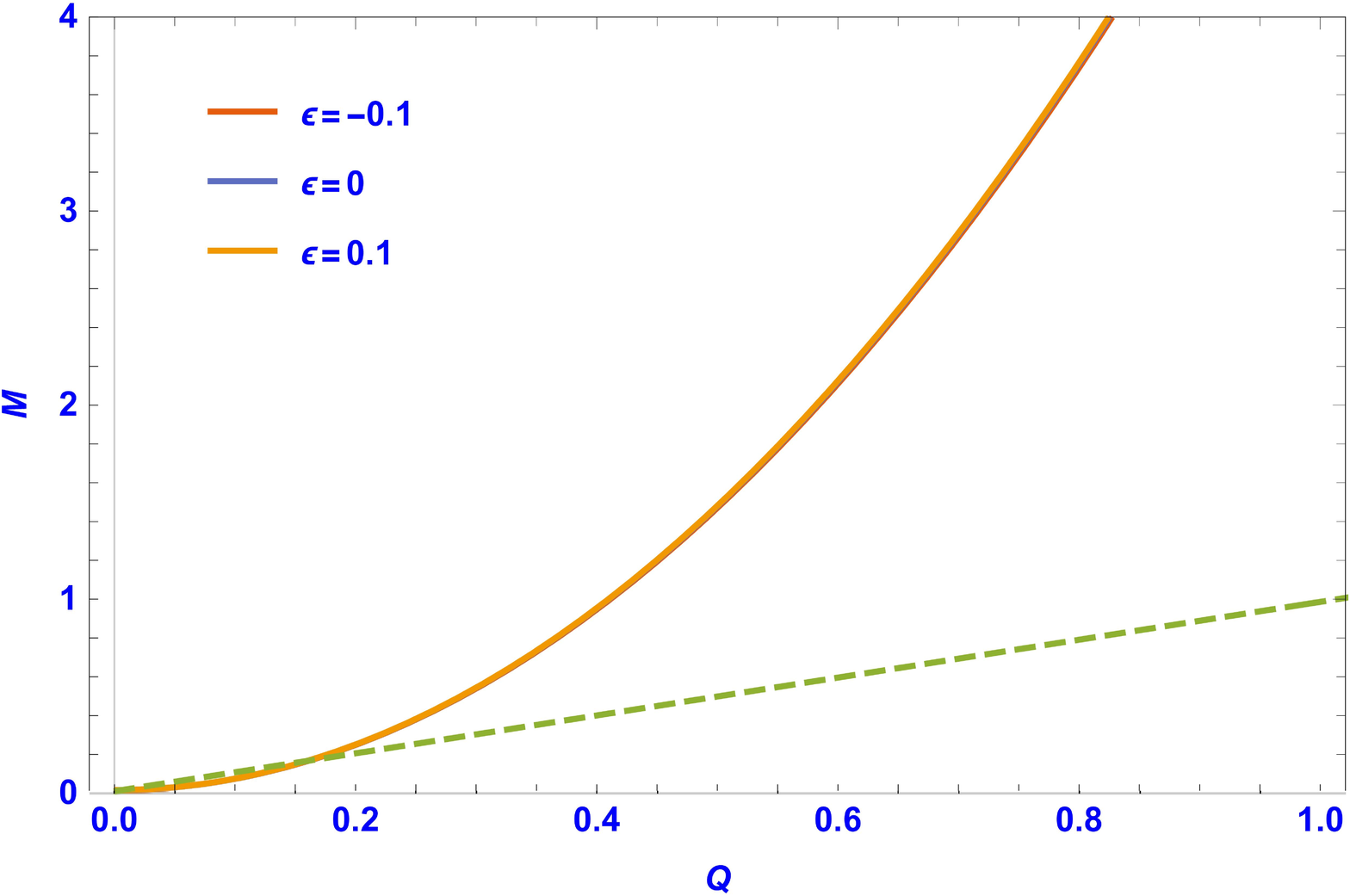}
\label{2b}}
\caption{\small{The plot of unmodified $M$ in term of $Q$ with respect to $l=0.1,0.2,0.3$ in  (a) and the plot of modified $M$ in term of $Q$ with respect to $l=0.1$, $\omega=\frac{-2}{3}$ and $\epsilon=-0.1,0,0.1$ in (b). }}
\label{fig2}
\end{center}
\end{figure}
According to the mass charge plot of the modified thermodynamic relations for the Kerr Newman AdS black hole as shown in Fig.
\ref{fig2}, we want to describe the concepts of WGC.
We compare the mass charge ratio and thermodynamic quantities of  modified and unmodified  cases to each other.   The charge and mass are equal in the extremal state   or the  mass charge ratio is unit as shown in the figure by dashed line. For the uncorrected mass mode, the mass-to-charge ratio is greater than one, we have plotted for different radius $l$ of AdS, as shown in the figure \ref{2a}. Now we  look at the modified state and use the corrected thermodynamic parameters.  We evaluate the amount of change in this correction as shown in figure (\ref{2b}. When the constant correction is negative, the mass of the black hole decreases, and when the small correction is positive, the mass increases. Therefore, when we consider small negative correction, the charge-to-mass ratio always increases. So, the negative  $\epsilon$  correction to the action show us this   black hole can behave like WGC. Thus, we confirm that the    corrections  play a very important role in the concept of WGC.
\subsection{Rotating Bardeen black holes in AdS space surrounded by perfect fluid}
In this section,   we check  universal relations for the rotating Bardeen black holes in AdS space surrounded by perfect fluid \cite{081}. We note here that   black holes have singularity \cite{073}and in such case  space-time, densities and curvatures tend to infinity \cite{074,075}.  The physical predictions in these points face serious problems and  these singularities always cause serious problems for general relativity. But Bardeen   black hole solution   includes a new description of black holes without singularity \cite{076, 077}.   Now,  in order  to investigate new universal relation  here we introduce the  magnetic charge $\vartheta$   to the theory and  see the effect of such parameter to universal relation.  The rotating Bardeen black holes in AdS space surrounded by perfect fluid  is described by
\begin{equation} \label{85}
ds^{2}=f(r)dt^{2}-f^{-1}(r)dr^{2}-r^{2}d\Omega^{2},
\end{equation}
with
\begin{equation}\label{86}
 f(r)=r^{2}+a^{2}-\frac{2M r^{4}}{(r^{2}+\vartheta^{2})^{\frac{3}{2}}}+\frac{r^{2}}{l^{2}}+\alpha r\ln\frac{r}{|\alpha|},
\end{equation}
where  $M$  $a$, $\alpha$ and $\vartheta$ denote    mass, rotational parameter, perfect fluid  parameter and magnetic charge parameter, respectively. With respect to entropy of black hole, we investigate the thermodynamic relation of rotating Bardeen black holes in AdS space surrounded by perfect fluid. Here, we consider a small constant correction $\epsilon$ and with respect to\cite{081}, we calculate the modified thermodynamic quantities as given by
\begin{equation}\label{87}
M=\frac{(S+\pi\vartheta^{2})^{\frac{3}{2}}}{2S^{2}\sqrt{\pi}}((a^{2}\pi+S)+\frac{S(1+\epsilon)}{l^{2}}+
\alpha\sqrt{S\pi}\log(\frac{\sqrt{S}}{\alpha\sqrt{\pi}}),
\end{equation}
\begin{equation}\label{88}
T=\frac{(S+\pi\vartheta^{2})^{\frac{3}{2}}}{2S^{2}\sqrt{\pi}}\left(-1-\frac{2a^{2}\pi}{S}+\frac{\sqrt{\pi}\alpha}{2\sqrt{S}}+\frac{3(a^{2}\pi+S)}{2(S+\pi\vartheta^{2})}+\frac{(S-2\pi\vartheta^{2})(1+\epsilon)}{2l^{2}(S+\pi\vartheta^{2})}-\frac{3\vartheta^{2}\alpha \pi^{\frac{3}{2}}\log(\frac{\sqrt{S}}{\sqrt{\pi}\alpha})}{2\sqrt{S}(S+\pi \vartheta^{2})}\right),
\end{equation}
and
\begin{equation}\label{89}
\Omega=\frac{a\sqrt{\pi}(S+\pi \vartheta^{2})^{\frac{3}{2}}}{S^{2}}.
\end{equation}
Due to thermodynamic relation,  we consider  $\eta=\frac{(S+\pi\vartheta^{2})^{\frac{3}{2}}(-1+\log(\frac{\sqrt{S}}{\alpha\sqrt{\pi}}))}{2S^{\frac{3}{2}}}$ conjugate to perfect fluid $\alpha$, volume $V=\frac{4\sqrt{\pi}(S+\pi \vartheta^{2})^{\frac{3}{2}}(-1+\log(\frac{\sqrt{S}}{\alpha\sqrt{\pi}}))}{3S}$ conjugate to $P=-\frac{\Lambda}{8\pi}$ and  $\zeta=\frac{3M\pi\vartheta}{S+\pi\vartheta^{2}}$  conjugate to $\vartheta$. To study the universal relation we solve the equation (\ref{87}). So the constant correction parameter$\epsilon$ is given by
\begin{equation}\label{90}
\epsilon=-1+\frac{2l^{2}MS\sqrt{\pi}}{(S+\pi\vartheta^{2})^{\frac{3}{2}}}-\frac{l^{2}(a^{2}\pi+S+\sqrt{\pi S}\alpha\log(\frac{\sqrt{S}}{\alpha\sqrt{\pi}}))}{S}.
\end{equation}
The derivative  of $\epsilon$ with respect to $S$ gives
\begin{equation}\label{91}
\frac{\partial\epsilon}{\partial S}=-\frac{l^{2}M \sqrt{\pi}(S-2\pi\vartheta^{2})}{(S+\pi\vartheta^{2})^{\frac{3}{2}}}+\frac{l^{2}\sqrt{\pi}(2a^{2}\sqrt{\pi}-\alpha\sqrt{S}+\alpha\sqrt{S}\log(\frac{\sqrt{S}}{\alpha\sqrt{\pi}}))}{2S^{2}}.
\end{equation}
We combine two equation (\ref{88}) and (\ref{91}), one can obtain
\begin{equation}\label{92}
-T\frac{\partial S}{\partial \epsilon}=\frac{(S+\pi\vartheta^{2})^{\frac{3}{2}}}{2l^{2}S\sqrt{\pi}}.
\end{equation}
By setting $ T = 0 $ and  using equations (\ref{87}) and (\ref{88}), we get
\begin{equation}\label{93}
\frac{\partial M_{ext}}{\partial\epsilon}=\frac{(S+\pi\vartheta^{2})^{\frac{3}{2}}}{2l^{2}S\sqrt{\pi}}.
\end{equation}
We see here  two equations (\ref{92}) and (\ref{93}) are exactly  same.

 Now,  we   use the pressure $P=\frac{3}{8\pi l^{2}}=-\frac{\Lambda}{8\pi}$ and equation (\ref{90}) to obtain
\begin{equation}\label{94}
\frac{\partial \epsilon}{\partial P}=\frac{3(\frac{1}{\pi}+\frac{a^{2}}{S}-\frac{2MS}{\sqrt{\pi}(S+\pi\vartheta^{2})^{\frac{3}{2}}}+\frac{\alpha\log(\frac{\sqrt{S}}{\alpha\sqrt{\pi}})}{\sqrt{\pi S}})}{8P^{2}}.
\end{equation}
According to volume and    extremal bound, we have
\begin{equation}\label{95}
-V\frac{\partial P}{\partial \epsilon}=\frac{(S+\pi\vartheta^{2})^{\frac{3}{2}}}{2l^{2}S\sqrt{\pi}}.
\end{equation}
Here  two equations (\ref{95}) and (\ref{93}) are exactly  same.

Also with respect to equation (\ref{90}), we get
\begin{equation}\label{96}
\frac{\partial \epsilon}{\partial a}=-\frac{2al^{2}\pi}{S}.
\end{equation}
 So from (\ref{89}) and (\ref{96}), we obtain following relation:
\begin{equation}\label{97}
-\Omega\frac{\partial a}{\partial \epsilon}=\frac{(S+\pi\vartheta^{2})^{\frac{3}{2}}}{2l^{2}S\sqrt{\pi}}.
\end{equation}
We see here, two equations (\ref{95}) and (\ref{97}) are exactly  same.

 We use equation (\ref{90}) for obtaining the another universal repletion. Here, we have
\begin{equation}\label{98}
\frac{\partial \epsilon}{\partial\alpha}=\frac{l^{2}\sqrt{\pi}}{\sqrt{S}}-\frac{l^{2}\sqrt{\pi}\log(\frac{\sqrt{S}}{\alpha\sqrt{\pi}})}{\sqrt{S}}.
\end{equation}
So by using the $\eta$ and (\ref{98}), we obtain the other universal relation as given by
\begin{equation}\label{99}
-\eta\frac{\partial \alpha}{\partial\epsilon}=\frac{(S+\pi\vartheta^{2})^{\frac{3}{2}}}{2l^{2}S\sqrt{\pi}}.
\end{equation}
Also here, two equation (\ref{99}) and (\ref{97}) are extremely same.

Again we use
relation (\ref{90}) and obtain
\begin{equation} \label{100}
\frac{\partial \epsilon}{\partial \vartheta}=-\frac{6l^{2}M\pi^{\frac{3}{2}}S
\vartheta}{(S+\pi\vartheta^{2})^{\frac{5}{2}}}.
\end{equation}
So, with respect the above equation and  $\zeta$, we obtain
the new universal relation of rotating Bardeen black hole as given by
\begin{equation}\label{101}
-\zeta\frac{\partial \vartheta}{\partial \epsilon}=\frac{(S+\pi
\vartheta^{2})^{\frac{3}{2}}}{2l^{2}S\sqrt{\pi}}=\frac{\partial M_{ext}}{\partial
\epsilon}.
\end{equation}
The most important thing here is that, the universal relation related to magnetic
charge is also confirmed.

We thoroughly investigated the universal relations for the three different black
holes in the AdS space surrounded by perfect fluid, such as Kerr-Newman, rotating
Bardeen and Reissner-Nordstr\"{o}m. We calculated the universal relationships of
each black hole separately. We also introduced new universal relations with
respect to the concepts of perfect fluid and string fluid. We also observed that
when a small correction constant is added to the action,   the modified
thermodynamic quantities and  relations can be
calculated. This constant correction can lead to a decrease of mass and  increase
the charge-to-mass ratio, which is a clue of WGC behaviour. The
concepts in this paper can be evaluated for other black holes with different
properties, as well as considering the higher dimensions and black holes in
different structures.
\section{Conclusions}
Researchers have acquired new universal relations from multiple methods for black holes in the last few years. These universal relations can be an excellent impetus for integrating different sciences and possibly a great solution to the path of quantum gravity for physicists. In this paper, we confirmed new universal relations for black hole thermodynamics. We investigated each of these universal relations by selecting different black holes such as AdS Schwarzschild, charged BTZ, charged rotating BTZ, accelerating and charged accelerating black holes, and AdS black hole surrounded by perfect fluid. First, we obtained the modified thermodynamic relations of the black holes assuming a small correction to the action. We confirmed the universal relations by performing a series of direct calculations. It is noteworthy that according to each of the properties related to black holes, such as rotating, charged, accelerating, etc., a new universal relation can be obtained according to this method. So, we have confirmed two different types of these universal relations for various block holes. One of the most valuable results is that using the unique feature of black holes, a new universal relation between different black holes thermodynamics can be investigated. For  Kerr-Newman, rotating Bardeen and Reissner-Nordstr\"{o}m black holes in the AdS space surrounded by perfect fluid we also consider a small constant correction to the action and computed modified thermodynamic quantities and relations. Then, by using a series of calculations, we obtained the universal relations of these black holes. We have investigated new universal relations related to the parameter of perfect fluid and magnetic charge. Also, according to the constant correction and universal relations, we have studied the effect of perfect fluid and magnetic parameters on the charge-to-mass ratio. Here, we have seen that the WGC-like behavior is satisfied by the black hole system. This work will also be interesting in investigating the universal relation of the black hole in higher dimensions. Also, it may interesting to do correction the Einstein-Gauss-Bonnet action and obtain the new modified thermodynamic relations. Given the relationship between the universal relation and the WGC, it may be interesting to obtain the relation between the correction parameter to the action and the Gauss-Bonnet parameter.\\

Conflict of Interest\\
The authors declare that they have no known competing financial interests or personal relationships that could have appeared to influence the work reported in this paper.\\
Data Availability Statement\\
Data sharing is not applicable to this article as no datasets were generated or analyzed during
the current study.\\

\end{document}